\documentclass[%
reprint,
superscriptaddress,
 amsmath,amssymb,
 aps,
floatfix,
]{revtex4-2}
\usepackage{mathbbol}
\usepackage{amssymb}   % AMS Math
\usepackage{comment}
\DeclareSymbolFontAlphabet{\amsmathbb}{AMSb}%
\usepackage{graphicx}% Include figure files
\usepackage{dcolumn}% Align table columns on decimal point
\usepackage{bm}% bold math
\usepackage{amsmath,amsfonts,amsthm, mathtools}
\usepackage{xcolor}
\usepackage{braket}
\usepackage{soul}
\DeclareMathOperator{\sinc}{sinc}

\theoremstyle{definition}

\usepackage{color}

\begin{document}

\preprint{APS/123-QED}

\title{Demonstrating Superresolution in Radar Range Estimation Using a Denoising Autoencoder }

%Authors - To be modified

\author{Robert Czupryniak}
\affiliation{Department of Physics and Astronomy, University of Rochester, Rochester, NY, 14627, USA}
\affiliation{Institute for Quantum Studies, Chapman University, Orange, CA, 92866, USA}

\author{Abhishek Chakraborty}
\affiliation{Department of Physics and Astronomy, University of Rochester, Rochester, NY, 14627, USA}
\affiliation{Institute for Quantum Studies, Chapman University, Orange, CA, 92866, USA}

% \author{...}
\author{Andrew N. Jordan}
\affiliation{Institute for Quantum Studies, Chapman University, Orange, CA, 92866, USA}
\affiliation{Schmid College of Science and Technology, Chapman University, Orange, CA, 92866, USA}
\affiliation{The Kennedy Chair in Physics, Chapman University, Orange, CA, 92866, USA}
\affiliation{Department of Physics and Astronomy, University of Rochester, Rochester, NY, 14627, USA}
\author{John C. Howell}
\affiliation{Institute for Quantum Studies, Chapman University, Orange, CA, 92866, USA}
\affiliation{Schmid College of Science and Technology, Chapman University, Orange, CA, 92866, USA}

\newcommand{\abhishek}[1]{\textcolor{violet}{#1}}

\begin{abstract}
We apply machine learning methods to demonstrate range superresolution in remote sensing radar detection.  Specifically, we implement a denoising autoencoder to estimate the distance between two equal intensity scatterers in the subwavelength regime. 
The machine learning models are trained on waveforms subject to a bandlimit constraint such that ranges much smaller than the inverse bandlimit are optimized in their precision. 
The autoencoder achieves effective dimensionality reduction, with the bottleneck layer exhibiting a strong and consistent correlation with the true scatterer separation.
We confirm reproducibility across different training sessions and network initializations by analyzing the scaled encoder outputs and their robustness to noise.
We investigate the behavior of the bottleneck layer for the following types of pulses: a traditional sinc pulse, a bandlimited triangle-type pulse, and a theoretically near-optimal pulse created from a spherical Bessel function basis. The Bessel signal performs best, followed by the triangle wave, with the sinc signal performing worst, highlighting the crucial role of signal design in the success of machine-learning-based range resolution.
 
\end{abstract}

\maketitle

\section{Introduction}
Recent discoveries in radar ranging \cite{wehner1987high, li2001moving, cooper2008high, zhang2017photonics, skolnik1980introduction, levanon2004radar} have shown that previous traditional bounds can be surpassed \cite{howell2023super}.  These previous bounds are based on the principle that in order to resolve two nearby scatterers, the temporal duration of the probing radar pulse must be smaller than the scatterer separation \cite{skolnik1980introduction,sheriff1977limitations,neal2004ground}.  While many applications in radar could greatly benefit from breaking this tradeoff – for example when the medium imposes a bandlimit on EM wave propagation - to go beyond this limit was thought to be impossible.  

The challenge is to make short distance range estimates with long wavelength waves. However, it was recently shown that even in the context of classical wave theory, this bound could be beaten by orders of magnitude by using tailored pulses with a self-referencing estimation theory approach \cite{howell2023super}.  

There are similarities to the breaking of the Raleigh criterion in classical optical imaging systems using mode sorting techniques \cite{PhysRevX.6.031033} - however, unlike those techniques, we deal with fully coherent wave physics and direct detection.  Subsequent work proved new bounds on this type of estimation technique using Fisher information methods by incorporating the technical noise added by the environment and amplifiers \cite{jordan2023fundamental}.  Optimal pulses were also derived using a bandlimit as a constraint and optimizing the Fisher information about the separation parameter, given prior knowledge of two scatterers \cite{jordan2024best}.

While theoretical results and examples have also been worked out for a multiparameter scatterer system \cite{jordan2023fundamental}, it is of great interest to go to a more model-agnostic type of estimation method.  As such, it is natural to consider machine learning (ML) approaches to this problem.  It is of interest to see to what extent the ML methods can rank pulse shapes and even find new pulse shapes in order to estimate the scatterer separation under different types of constraints, as well as compare its performance with the previously mentioned fundamental bounds.  While more general types of target recognition will be considered in the future, we focus here on the simplest case of balanced reflectors as a test bed for these methods.

\begin{figure}
    \centering
    \includegraphics[width=\columnwidth]{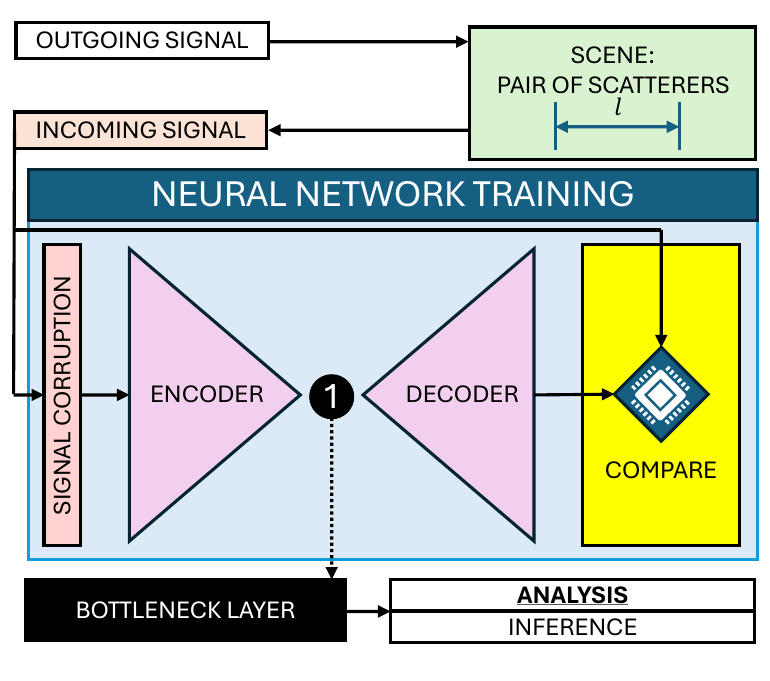}
    \caption{
    Schematic overview of the autoencoder training process. 
    The scene consists of a pair of scatterers, generating an incoming signal that is deliberately corrupted with white Gaussian noise before being processed by the neural network. The neural network's primary goal is to denoise the input and accurately reconstruct the original signal.
    The neural network under investigation is an autoencoder, which consists of two main components: an encoder and a decoder. The encoder learns a compressed representation of the input signal, referred to as the \textit{bottleneck layer} (labeled, 1 neuron wide), while the decoder takes this compressed representation as input and attempts to reconstruct the original, noise-free signal. The objective of this paper is to examine the behavior of the bottleneck layer in trained autoencoders, with a focus on different types of outgoing signals.}
    \label{fig:overview}
\end{figure}

It was recently shown that self-referencing waves, with local structure that can be distinguished using interferometric imaging, can provide subwavelength resolution below the Rayleigh limit \cite{howell2023super,jordan2023fundamental}. This technique shows promise for sub-wavelength imaging, at the cost of a numerically intensive estimation algorithm due to the large dimensionality of the return signal. For complex signals, a linear dimensionality reduction technique such as principal component analysis (PCA) \cite{Jolliffe2002Principalcomponentanalysis} might be suboptimal for dimensionality reduction/data compression. Hence, this problem is well suited for machine-learning based techniques.

Artificial neural networks (ANN) and deep-learning based approaches have shown great success in the field of image processing and computer vision, to the extent that most leading algorithms to solve the ImageNet Large Scale Visual Recognition Challenge (ILSVRC) \cite{ImageNet} were based on deep-learning approaches. In particular, AlexNet \cite{alexnet}, a deep convolutional neural network, surpassed  previous heurisitic image-processing techniques based on manually designed filters for detection of various features in images. Subsequently, a much deeper (152 layers) network (called ResNet \cite{imagenet-resnet-ms}) featuring residual connections performed even better. 

The success of these techniques can be attributed to the fact that the deep-learning techniques effectively offer an automated process to design feature-detection filters based on the training data. However, it becomes increasingly more difficult to attribute such a simple working principle to the success of more complex and deeper ANNs, thus making them seem like black boxes. 

Representation learning \cite{bengio2014representationlearningreviewnew}, which aims to learn a lower dimensional representation of much higher dimensional data, has been used successfully for many tasks, including object detection \cite{kaur2023comprehensive}, image segmentation \cite{image_segmentation} and super-resolution imaging \cite{SR_photo_1, ESRGAN}. Such representations can be learned using a particular ANN architecture called autoencoders \cite{autoencoders}, which use an encoder-decoder structure with a bottleneck layer in between to learn a better representation of the data. 

Autoencoders allow for a semi-supervised/unsupervised learning approach to this problem, where the neural networks are trained to reconstruct the clean lower-dimensional representation of (potentially noisy) data, thus eliminating the need to manually assigned labels for classification/regression tasks. Instead the representation learned in the bottleneck layer can be used as a label for further tasks downstream in the data-processing pipeline.

In this article, we examine/demonstrate the use of a denoising autoencoder \cite{denoising_autoencoders} (DAE) to solve the single-parameter radar range resolution problem for a symmetric two-scatterer system. Leveraging the autoencoder, we evaluate the performance of three distinct radar pulse candidates within an unsupervised learning framework. Specifically, we analyze the autoencoder’s denoising capabilities for various incoming, bandlimited signals. The signals we consider were designed to resolve separations below the inverse bandwidth of the signal; hence, we focus on that regime. For this problem, we show that the bottleneck layer activation is strongly correlated with the scatterer separation, thus demonstrating successful dimensionality reduction. Our technique is general and can be applied to other parameter estimation problems or candidate signals without any significant changes. 

In Sec.~\ref{sec:range-resolution}, we review the radar range resolution problem. Sec.~\ref{sec:autoencoders} describes our deep-learning architecture and estimation procedure. Sec.~\ref{sec:signals} reviews the three different candidate waves we compare for range resolution. Finally, in Sec.~\ref{sec:results}, we describe our results. 

\section{Range resolution}\label{sec:range-resolution}

Radar ranging is a special case of imaging using radio waves. In a one-dimensional setting (such as is valid in the far-field limit), given an object or scene represented by a function $O(x)$, which is illuminated using a point spread function $f(x)$, the image is given by a convolution of $O(x)$ and $f(x)$

\begin{equation}
    I(x) = O(x) \ast f(x) = \int_{-\infty}^{\infty} O(y) f(y-x) dy.
    \label{eq:imaging-convolution}
\end{equation}

We consider the simple problem of estimating the separation between two equal magnitude point scatterers in range resolution. The object function is represented by two delta-function spikes separated by a distance $l$. Let $f_0(x)$ be the outgoing signal to probe the scatterer separation $l$, the return signal has the form
\begin{equation}
\begin{split}
    f^{(l)}(x) &= \left[f_0(x-l/2) + f_0(x+l/2) \right]/2+\epsilon(x) \\
    &= f_c(x) + \epsilon(x),
\end{split}
\label{eq:two-scatterer-incoming-signal}
\end{equation}
where for convenience we set the origin at $l=0$, $\epsilon(x)$ is the noise, and $f_c(x)$ denotes the noiseless return signal. $f^{(l)}(x)$ is then studied with the goal of estimating $l$.
All of these signals are band-limited and exist in finite range of frequencies $f\in[0,F]$, establishing a trade-off between the cut-off frequency and range-resolution abilities.

\section{Autoencoders}\label{sec:autoencoders}

For this article, an autoencoder \cite{autoencoders} is defined by a pair of functionals, $\mathcal{E}: \mathbb{R}^N \rightarrow \mathbb{R}^n$ and $\mathcal{D}: \mathbb{R}^n \rightarrow \mathbb{R}^N$, both represented by neural networks. The input and output of this architecture are both vectors $\mathbf{X} \in \mathbb{R}^{N}$. The output of the encoder is typically labeled $\mathbf{z} = \mathcal{E}[\mathbf{X}] \in \mathbb{R}^{n}$. For undercomplete autoencoders, $N > n$, related to a compression or dimensionality reduction of the data. This architecture can be trained in an unsupervised manner by minimizing the reconstruction error,  given by the Euclidean distance (in $\mathbb{R}^{N}$) between the input and output which is equal to the $L_2$ norm $\lVert \mathcal{D}\left[ \mathcal{E}[\mathbf{X}] \right] - \mathbf{X} \rVert_2$. Once trained, the network is thought to have learned a compressed $n$-dimensional representation of the $N$-dimensional data, distilling the most important information about the input vector $\mathbf{X}$ into the compressed vector $\mathbf{z}$. Since artificial neural networks are universal function approximators, this can be thought of as a generalized nonlinear version of principal component analysis \cite{Jolliffe2002Principalcomponentanalysis}. 

\subsection{Denoising Autoencoders}

While regular autoencoders can effectively reconstruct noiseless signals for a variety of cases, the architecture can be adapted for denoising \cite{deeplearningbook, denoising_autoencoders} tasks as well, leading to the so-called denoising autoencoder. In this variant, the encoder is fed a noisy input vector $\tilde{\mathbf{X}}$, obtained by adding a zero-mean noise vector $\boldsymbol{\xi} 
= \mathcal{N}_N(\mathbf{0}, \boldsymbol{\sigma}_\xi)$ to the ``clean'' vector $\mathbf{X}$. We use $\mathcal{N}_N(\mathbf{0},  \boldsymbol{\sigma}_\xi)$ to denote a vector with $N$ elements, each of which is sampled from an independent, univariate normal distribution $\mathcal{N}(\mathbf{0},  \boldsymbol{\sigma}_\xi)$ with zero mean and $ \boldsymbol{\sigma}_\xi$ standard deviation. The resulting noisy signal vector is

\begin{equation}
    \tilde{\mathbf{X}} = \mathbf{X} + \boldsymbol{\xi}.
    \label{eq:noisy-vector-construction}
\end{equation}
The denoising autoencoder  is then trained to reconstruct, from the noisy data, the clean vectors $\mathbf{X}$ by minimizing the modified reconstruction loss $\lVert \mathcal{D}[ \mathcal{E}[\tilde{\mathbf{X}}] ] - \mathbf{X} \rVert_2$. This architecture has been used for denoising images with good results \cite{denoising_autoencoders}.
While the techniques described below still apply for different kinds of noise, an extensive comparison of the effect of different noise spectra is beyond the scope of this article. 

In the experimental setup, this noise model corresponds to adding white Gaussian detector noise to the incoming signal, similarly to the models considered in Refs.~\cite{howell2023super} and \cite{jordan2024best}. A more realistic model would include noise in the incoming signal arising from its propagation to the target scene and back to the detector. We leave this consideration beyond the scope of the present paper. 

\subsection{Autoencoder for Radar Range Resolution}\label{sec:ae_for_radars}

In the remainder of this article, we will focus on the problem described in Sec.~\ref{sec:range-resolution}. We consider a one-dimensional compressed representation $(n=1)$ since we are only concerned with a single parameter estimation problem. For this setup, we expect that, once trained, the activation of the single bottleneck layer neuron $z$ will be correlated with the separation parameter $l$. First, we examine this setup for noiseless signals to obtain the best case scenario. Next, we corrupt the input signal with noise and train the denoising autoencoder to reconstruct the noiseless signal. We will compare the information learned by the bottleneck layer neuron in both cases.

For a given outgoing signal $f_0(x)$ and scatterer separation $l$, the noiseless incoming signal $f^{(l)}(x)$ is generated numerically using Eq.~(\ref{eq:two-scatterer-incoming-signal}) and (optionally) corrupted by adding noise. The noiseless, noisy signals are then sampled and encoded in the vectors $\tilde{\mathbf{X}}^{(l)}, \mathbf{X}^{(l)}\in \mathbb{R}^{N}$ respectively. As a result, one can perform the estimation $l$ either using the large vector $\mathbf{X}^{(l)}$ representing the return signal, or the much smaller latent representation $\mathbf{z}^{(l)} = \mathcal{E}[\mathbf{X}^{(l)}]$.
For instance, the trained encoder output can be used as input to another numerical or neural model to estimate $l$. Since the dimension of $\mathbf{z}^{(l)}$ is much smaller than the dimension of $\mathbf{X}^{(l)}$, the estimation process should be computationally less intensive when performed on the encoder output rather than the sampled signal vector. We focus solely on the analysis of bottleneck layer $\mathbf{z}^{(l)}$ activation, showing that it correlates with the separation $l$; An extensive analysis of parameter estimation by inverting this relationship is beyond the scope of this paper. 

\section{Outgoing signals}\label{sec:signals}
We will measure lengths in units of the inverse bandwidth $1/F$. We select three types of band-limited outgoing signals for this analysis. First, we consider the $\sinc$ function raised to the $m^{\rm th}$ power
\begin{equation} \label{eq:f_S}
    f_S(t) = \sinc\left(\frac{\omega t}{ m \pi}\right) ^ m,
\end{equation}
taken for $m=10$ and $\omega=2\pi$ and scaled to give bandwidth $F$. This function was considered as a \textit{canvas} function to prepare superoscillating signals \cite{soda2020superoscillations}. The second is the band-limited \textit{triangle signal} defined in \cite{howell2023super}
\begin{equation} \label{eq:f_T}
    f_T(t) = p(t) f_S(t),
\end{equation}
which takes $f_S(t)$ as an envelope and multiplies it by the polynomial
\begin{equation} \label{eq:polynomial}
    p(t) = 8t - 14.3984 t^3 + 4.77612 t^5 - 0.82315 t^7.
\end{equation}
While $f_T$ should outperform $f_S$ when used in single parameter estimation \cite{howell2023super}, it is not the optimal signal for that purpose.

\begin{figure}
    \centering
    \includegraphics[width=\columnwidth]{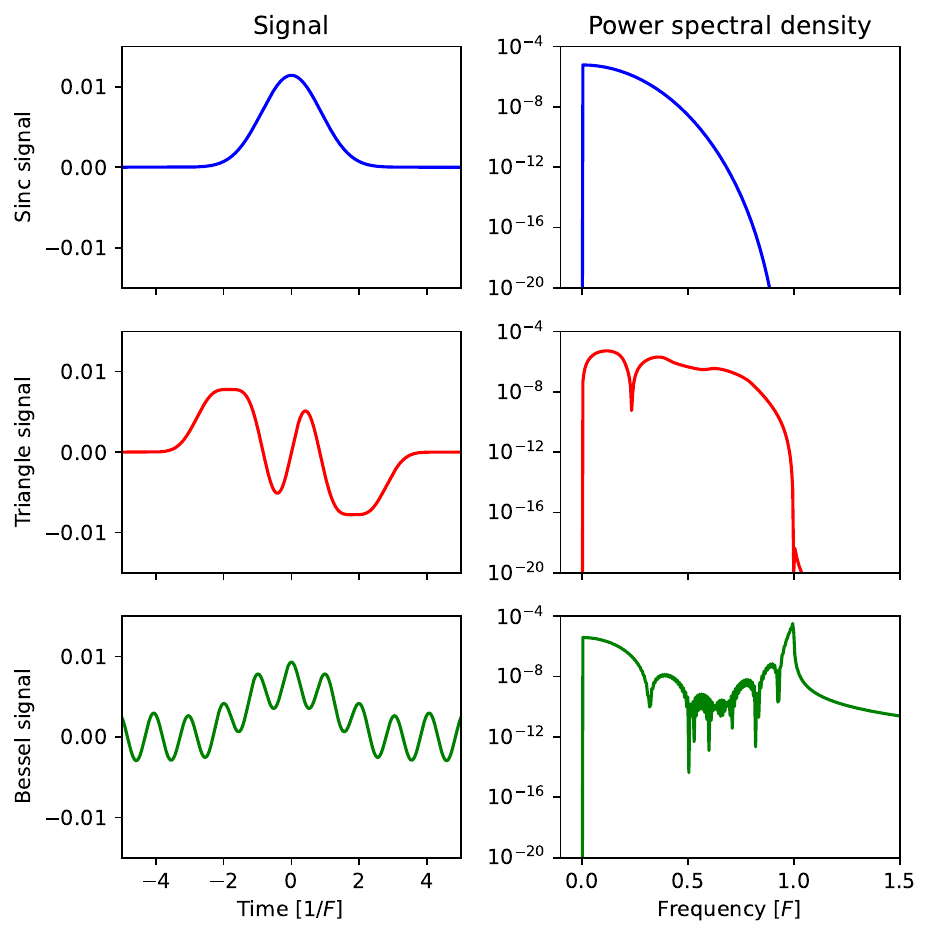}
    \caption{
    Outgoing signals (left) and their corresponding power spectra (right), used to estimate scatterer separation. The three signals shown are defined in Eqs.~(\ref{eq:f_S}), (\ref{eq:f_T}) and (\ref{eq:f_B}): the sinc pulse, the bandlimited triangle pulse, and the Bessel-based pulse. The high-frequency tail in the power spectrum of the Bessel signal arises from the finite sampling window, which is truncated at \( t = \pm 5 \, [1/F] \).}
    \label{fig:outgoing_signal}    
\end{figure}

An approach to find such optimal signals based on orthogonal function theory was proposed in Ref. \cite{karmakar2023beyond} and studied in Ref.~\cite{jordan2024best}, from which we take the following signal
\begin{equation} \label{eq:f_B}
\begin{aligned}
    f_B(t) & = 0.259858 \ j_0(\omega t) + 0.0879936 \ j_2(\omega t)  \\ 
           & + 1.13614 \ j_4 (\omega t) - 0.136663 \ j_6(\omega t) \\
           & + 1.23652 \ j_8(\omega t) - 0.185957 \ j_{10}(\omega t) \\
           & + 0.565418 j_{12}(\omega t).
\end{aligned}
\end{equation}
$j_n(\cdot)$ denotes the $n^{\rm th}$ spherical Bessel functions \cite{arfkenmathmethods} taken for $\omega=2\pi$. We plot the signals and their power spectrum in Fig.~\ref{fig:outgoing_signal}. We note that $f_B(t)$ is optimal, in principle, for an infinitely long waveform. Practically, we sample the incoming signal over a finite range and $f_B(t)$ will have a significant fraction of its power beyond it. Nevertheless, it is still expected to outperform the other two signals considered here.

The incoming signals are studied for scatterer separations ranging from $0$ to $1/F$ and sampled in the range $\left[-5,5\right] 1/F $, for the sub-resolved case. We downsample to $N=1024$ samples from each signal and form the vector $\mathbf{X}^{(l)}$ out of it that is passed through the neural networks. The uncorrupted signals are normalized to $\lVert\mathbf{X}^{(l)}\rVert_2=1$ before entering the neural networks.

The neural networks are trained separately for each signal type---sinc ($ f_S $), triangle ($ f_T $), and Bessel ($ f_B $)---and for each considered signal-to-noise ratio (SNR): noiseless, $ 1/2 $, $ 2/3 $, $ 1 $, and $ 2 $. We calculate the SNR using the formula $ \lVert \boldsymbol{\xi} \rVert_2 / \lVert \mathbf{X}^{(0)} \rVert_2 $, where $ \lVert \cdot \rVert_2 $ denotes the $ L_2 $ norm. To avoid the SNR’s dependence on the separation parameter $ l $, we use the norm of the outgoing signal $ \lVert \mathbf{X}^{(0)} \rVert_2 $, rather than that of the incoming signal $ \lVert \mathbf{X}^{(l)} \rVert_2 $.

All networks are initialized with random weights and biases. For networks trained on noisy signals, a new set of noise vectors $ \boldsymbol{\xi} $ is generated in each training epoch and added to the signals. This procedure introduces inherent randomness into the optimization process. To ensure that our results are reproducible and not artifacts of random initialization or random noise generation, we independently train 20 identical autoencoder networks for each signal type and each signal-to-noise ratio.

\section{Results}\label{sec:results}

\subsection{Noiseless signal analysis}\label{sec:noiseless_signal}
Due to inherent randomness of neural network training, a given signal can result in different bottleneck layer representations when forward-passed through different instances of the trained neural networks, even when all networks have the the same architecture and we use the same training procedure. To make the results comparable across different networks, we first rescale the trained encoder output as follows.

\begin{figure}
    \centering
    \includegraphics[width=\columnwidth]{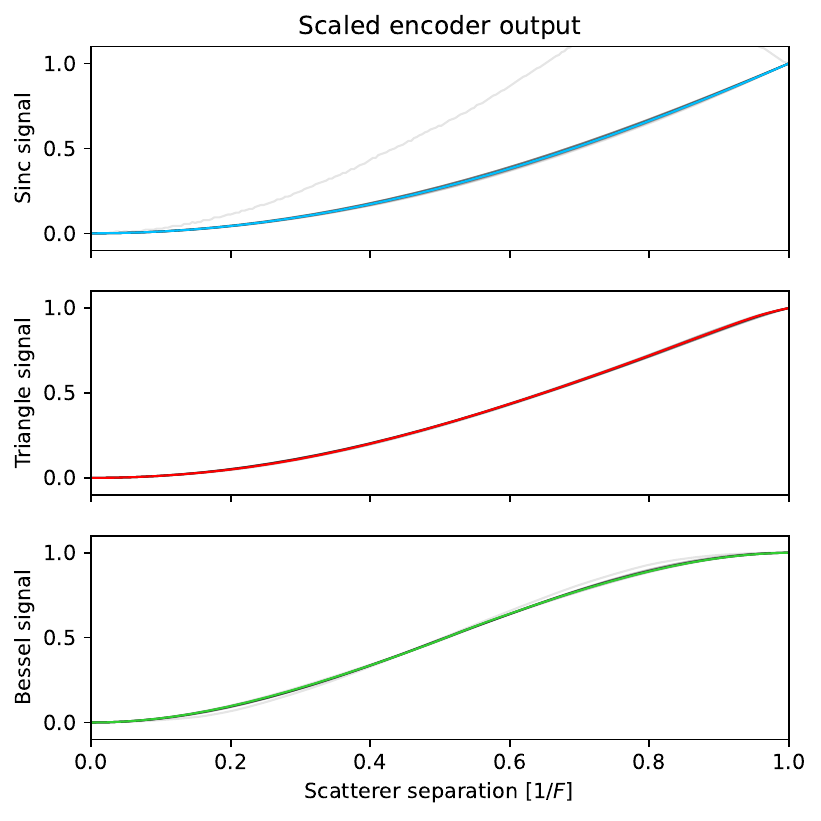}

    \caption{
    Scaled encoder output (as defined in Eq.~(\ref{eq:clean_signal_scaling})) plotted as a function of scatterer separation for each outgoing signal type. Each panel shows results from 20 independently trained neural networks, represented by gray curves. The colored curve overlaid on each plot corresponds to a representative single network. The consistent, monotonically increasing behavior across networks indicates reliable and reproducible learning of scatterer separation by the autoencoder bottleneck layer. Note that the top plot contains one curve not matching the rest, indicating unsuccessful training. 
    }
    \label{fig:clean_signal_encoder_output}    
\end{figure}

We generate a test dataset of $256$ incoming signals with scatterer separations between $0$ and $1/F$. For each signal we record the encoder output, resulting in the vector of encoder outputs $\mathbf{y}$ with entries $y_l = \mathcal{E}[\mathbf{X}^{(l)}]$. Let $y_i$ and $y_f$ be the initial and the final entry of $\mathbf{y}$. We obtain the scaled vector of encoder outputs from
\begin{equation}\label{eq:clean_signal_scaling}
    \tilde{\mathbf{y}} = \frac{\mathbf{y} - y_i}{y_f - y_i},
\end{equation}
which fixes the endpoints at $\tilde{y}_i = 0$ and $\tilde{y}_f = 1$. This scales the output of the final layer in a similar manner as batch normalization layer \cite{batchnorm}. The results are plotted in Fig.~\ref{fig:clean_signal_encoder_output}.

We observe that the curves representing the scaled encoder output as a function of scatterer separation have a roughly similar shape across all trained networks (see Fig.~\ref{fig:clean_signal_encoder_output}), indicating that the results are reproducible in different training sessions. All curves are monotonically increasing, making them suitable for parameter estimation. For each of them one should be able to define a mapping from scaled encoder output to scatterer separation resulting in near ideal estimation of $l$. We note that these statements hold even for a sinc signal (blue curves in Fig.~\ref{fig:clean_signal_encoder_output}) indicating that range resolution significantly below inverse bandwidth represents a simple task for neural networks in case of noiseless signals. Even though all three candidate signals are expected to perform similarly in the absence of noise, the advantage of $f_T$ and $f_B$ over $f_S$ becomes evident for the corrupted signals. 

Remarkably, the autoencoders develop meaningful representations in the bottleneck layer despite not being provided with scatterer separation labels during training. Furthermore, assuming access to the outgoing signal, one can identify the label corresponding to $l = 0$ in the noiseless model via Eq.~(\ref{eq:two-scatterer-incoming-signal}), since in this case the incoming and outgoing signals are identical. This enables qualitative comparisons across scenes, even without explicit knowledge of the underlying scatterer separations.

As an illustrative example, consider the task of comparing scatterer separations $l_1$ and $l_2$ in two distinct scenes. A forward pass of the outgoing signal through the encoder yields the reference output $y_0 = \mathcal{E}\left[\mathbf{X}^{(0)}\right]$, while the forward passes of the incoming signals yield $y_{l_{1(2)}} = \mathcal{E}\left[\mathbf{X}^{\left(l_{1(2)}\right)}\right]$. Comparing the magnitudes $|y_{l_1}-y_0|$ and $|y_{l_2}-y_0|$ allows one to infer which scene corresponds to the greater separation.

Quantitative estimation of scatterer separation, however, is more challenging—if not infeasible—as it would require knowledge of the encoder output as an explicit function of separation.

\subsection{Noisy signal analysis}\label{sec:noisy_signals}

In this section, we study the bottleneck layer of the denoising autoencoders trained to reconstruct a noiseless return signal from a noisy one. Note that each trained autoencoder corresponds to a specific candidate signal and SNR. For each network, we generate a dataset of $101$ noiseless signals, each with a different scatterer separation ranging from $l=0$ to $l=1[1/F]$. For each value of scatterer separation, we generate $1,000$ corrupted signals, each of them with a different realizations of noise. These $1,000$ signals are then forward passed through the autoencoder to produce as many encoder outputs. We then compute the mean and standard deviation of these values. The procedure is repeated for all values of scatterer separation. 

\begin{figure}
    \centering
    \includegraphics[width=\columnwidth]{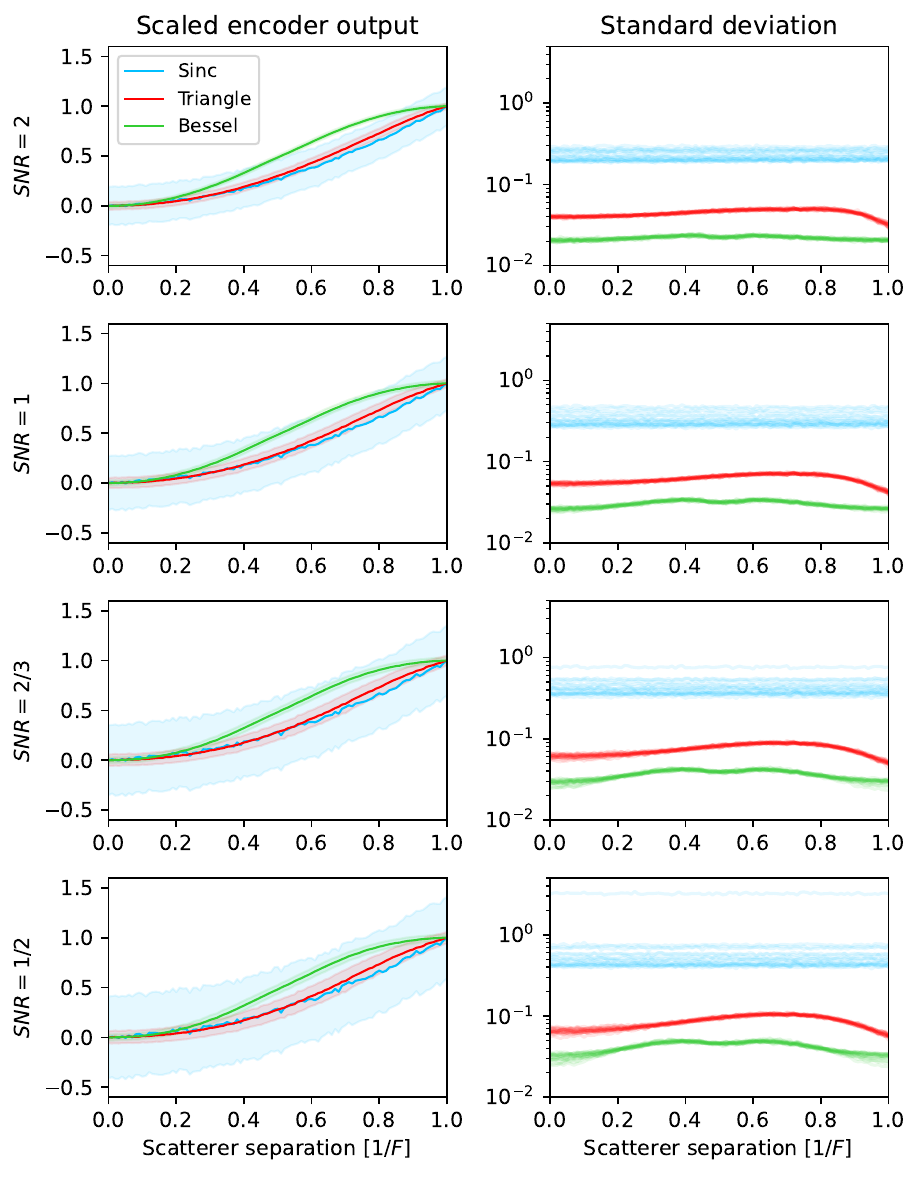}
    \caption{
    Left column: Scaled mean encoder output (solid lines), along with ± one standard deviation (shaded region), plotted as a function of scatterer separation. The scaling is performed according to Eq.~(\ref{eq:noisy_signal_scaling}). 
    For each separation value, $1,000$ noisy input signals were generated and passed through the autoencoder, and the mean and standard deviation of the scaled encoder output were computed.
    Each color represents a different type of incoming signal for a randomly selected autoencoder. 
    Right column: Standard deviation of the scaled encoder output plotted for all trained networks ($20$ curves per signal type). The uncertainty is highest for the sinc signal, intermediate for the triangle signal, and lowest for the Bessel-based signal. 
    Each row corresponds to autoencoders trained at a different signal-to-noise ratio (SNR).
    }
    \label{fig:noise_analysis}    
\end{figure}

As a result, we obtain a vector of means $\mathbf{m}$ and standard deviations $\boldsymbol{\sigma}$ with entries corresponding to computation results for a given scatterer separation, i.e.,  $m_k = m(l_k)$, $\sigma_k = \sigma(l_k)$, in increasing order of scatterer separation $l_k$. Due to inherent randomness of deep learning these vectors will be different for each neural network. Similar to noiseless signal analysis, we rescale the vectors $\mathbf{m}$ and $\boldsymbol{\sigma}$ to make the results comparable across different neural networks. Let's denote by $i$ and $f$ the initial and final elements of these vectors. We obtain the scaled vectors from
\begin{equation} \label{eq:noisy_signal_scaling}
    \tilde{\mathbf{m}} = \frac{\mathbf{m} - m_i}{m_f - m_i}, \quad 
    \tilde{\boldsymbol{\sigma}} = \left| (m_f - m_i)^{-1} \right|\boldsymbol{\sigma}.
\end{equation}
$\tilde{\boldsymbol{\sigma}}$ can be understood as the spread of the scaled encoder output due to noise in the input, and it can be treated as a measure of how well the corruption process is reversed by the encoder. The results produced by the described analysis are given in Fig.~\ref{fig:noise_analysis}.

\subsection{Discussion}

Observe that improving the quality of the outgoing signal leads to a substantial enhancement in the autoencoder’s ability to reverse the corruption process. Across all considered signal-to-noise ratio values, the sinc signal, $f_S(t)$, exhibited the poorest performance, corresponding to the highest noise levels in the bottleneck layer. This result is expected, as $f_S(t)$ lacks distinguishing features that would enable it to resolve separations below the inverse bandwidth.

In contrast, the triangular signal, $f_T(t)$, introduced in Ref.~\cite{howell2023super} can be considered as an ansatz suitable for estimating low scatterer separations. Indeed, autoencoders trained on incoming signals generated by $f_T(t)$ demonstrated significantly improved performance compared to those trained on sinc signals, as indicated by the red curves in Fig.~\ref{fig:noise_analysis}.

The Bessel signal, $f_B(t)$, studied in Refs.~\cite{jordan2024best, karmakar2023beyond}, was explicitly designed for estimating small scatterer separations. Consequently, it was expected to outperform the other signals in this task. This expectation was confirmed, as autoencoders trained on incoming signals from $f_B(t)$ achieved the highest efficiency in noise reduction within the bottleneck layer.

The findings presented in this section suggest a hierarchical ranking of outgoing signals based on their effectiveness in estimating small scatterer separations. The sinc signal, $f_S(t)$, lacks any structural features beneficial for this task. The triangular signal, $f_T(t)$, serves as an ansatz that is anticipated to yield better results, while the Bessel signal, $f_B(t)$, is grounded in solid theoretical principles and is expected to deliver the best performance. This hierarchy is reflected in the performance of the trained autoencoders, as illustrated in the right column of Fig.~\ref{fig:noise_analysis}.

It is crucial to emphasize that the only information provided to the autoencoders during training was the incoming signals. The autoencoders received no direct information about scatterer separation or signal-to-noise ratios. Remarkably, despite this limitation, the bottleneck layer consistently exhibited a structured response as a function of scatterer separation for both clean (Fig.\ref{fig:clean_signal_encoder_output}) and corrupted signals (Fig.\ref{fig:noise_analysis}). Furthermore, the noise levels in the bottleneck layer remained consistent across different neural networks trained with the same outgoing signals and noise levels, indicating that the stochastic nature of neural network training does not compromise our conclusions. In the Supplementary Material, we demonstrate that these observations hold even when different encoder architectures are employed.

\section{Conclusions and outlook}
We have demonstrated a proof-of-concept that autoencoders (specifically denoising autoencoders) can be used for parameter estimation in the radar range resolution problem. 

The results of our ML simulations showed that the bottleneck layer of the autoencoder develops a useful representation for the parameter estimation problem. If the network is trained on clean signals and one has access to scatterer separations as labels, it is to possible establish a relationship of the encoder output as a function of scatterer separation, enabling quantitative estimates. Even without access to labels, the bottleneck layer still permits qualitative analysis of different scenes, for example, allowing comparison of scatterer separations between scenes.

Interestingly, different autoencoders trained on a given signal develop similar representations in the bottleneck layer, despite the inherent randomness of the machine learning method. This similarity appears not only in the shape of the encoder output as a function of scatterer separation, but also in how the encoder output responds to noise in the incoming signal. These findings are consistent across different encoder architectures, as demonstrated in the Supplementary Material.

We have verified how the bottleneck layer is affected by noise depending on the incoming signal. As expected, the network exhibits less noise as we increase the quality of the incoming signal. To verify it, we compared three different signals: one based on a sinc wave to the $m^{\rm th}$ power, a band-limited \textit{triangle wave} and a signal constructed from Bessel functions, which is theoretically near-optimal for the single parameter problem examined here. The Bessel signal performs best, followed by the triangle wave, with the sinc signal performing worst. Our results demonstrate that improving the quality of the outgoing signal enhances the correlation between bottleneck layer activations and the scatterer separation in autoencoder-based analysis. This suggests that in addressing range resolution beyond the inverse bandwidth, optimizing neural network architectures alone is insufficient; poorly designed outgoing signals can significantly degrade network performance.

\section{Code availability statement}

The source code used to generate the results presented in this manuscript will be made publicly available upon acceptance in a peer-reviewed journal.

\section{Acknowledgments}
This work was supported by the Air Force Office of Scientific Research under Award No. FA9550-24-1-0329. We thank Achim Kempf, Barbara {\v S}oda, Nick Vamivakas, Shunxing Zhang and Derek White for helpful discussions.

\bibliographystyle{unsrt}
\bibliography{references}% Produces the bibliography via BibTeX.

\appendix

\section{Autoencoder architecture} \label{app:architecture}

The architecture described in Table~\ref{tab:cnn_ae} is used for all neural networks responsible for generating the results presented in the main body of the paper. In the Supplementary Material, we explore two additional encoder variants for comparison.

The first alternative encoder consists of a sequence of fully connected (linear) layers, as detailed in Table~\ref{tab:linear_encoder}. This architecture preserves the same decoder structure used in the main model.

The second alternative encoder begins with a custom Fourier transform layer followed by a low-pass filter. The input is a 1024-element vector representing the sampled signal. We compute its Fourier transform and retain only the frequency components below the inverse bandwidth. For the signal parameters considered in this work, this yields 11 frequency components. We extract both the real and imaginary parts of these components and concatenate them, producing a 22-element feature vector as the layer's output.

This 22-dimensional representation is then passed through a series of linear layers, forming the full encoder architecture described in Table~\ref{tab:fourier_encoder}. The same decoder is used as in the main architecture.

All networks were implemented and trained using PyTorch v2.7.0.

\begin{table*}[]
\resizebox{300pt}{!}{%
\begin{tabular}{l|l|l|}
\cline{2-3}
 & \textbf{Layer} & \textbf{Output size} \\ \hline
\multicolumn{1}{|l|}{\textbf{Encoder}} & Conv1d, input channels = 32, output channels = 32, kernel size = 64 & $(B,1,32,961)$ \\ \hline
\textbf{} & MaxPool1d, kernel size = 4 + GELU & $(B,1,32,240)$ \\ \cline{2-3} 
\textbf{} & Conv1d, input channels = 32, output channels = 32, kernel size = 32 & $(B,1,32,209)$ \\ \cline{2-3} 
\textbf{} & MaxPool1d, kernel size = 4 + GELU & $(B,1,32,52)$ \\ \cline{2-3} 
\textbf{} & Conv1d, input channels = 32, output channels = 32, kernel size = 16 & $(B,1,32,37)$ \\ \cline{2-3} 
\textbf{} & MaxPool1d, kernel size = 4 + GELU & $(B,1,32,9)$ \\ \cline{2-3} 
\textbf{} & Conv1d, input channels = 32, output channels = 32, kernel size = 9 & $(B,1,32,1)$ \\ \cline{2-3} 
\textbf{} & Flatten + GELU & $(B,32)$ \\ \cline{2-3} 
\textbf{} & Linear & $(B,1)$ \\ \hline
\multicolumn{1}{|l|}{\textbf{Decoder}} & Linear + Tanh & $(B,256)$  \\ \hline
\textbf{} & Linear  & $(B,1024)$ \\ \cline{2-3} 
\textbf{} & Reshape & $(B,1,1024)$ \\ \cline{2-3} 
\end{tabular}%
}
\caption{
Architecture of the autoencoder used in the main body of the paper. The encoder consists of convolutional layers followed by a fully connected layer, and the decoder consists of two fully connected layers. $B$ denotes the batch size.
}
\label{tab:cnn_ae}
\end{table*}

\begin{table}[]
\resizebox{100pt}{!}{%
\begin{tabular}{|l|l|}
\hline
\textbf{Layer} & \textbf{Output size} \\ \hline
Flatten & $(B,1024)$ \\ \hline
Linear + GELU & $(B,256)$ \\ \hline
Linear + GELU & $(B,64)$ \\ \hline
Linear + GELU & $(B,8)$ \\ \hline
Linear & $(B,1)$ \\ \hline
\end{tabular}%
}
\caption{The architecture of the linear encoder. The same decoder architecture is used as in Tab.~\ref{tab:cnn_ae}. $B$ indicates the batch size.}
\label{tab:linear_encoder}
\end{table}

\begin{table}[]
\resizebox{160pt}{!}{%
\begin{tabular}{|l|l|}
\hline
\textbf{Layer} & \textbf{Output size} \\ \hline
Fourier transform and low pass filter& $(B,1,22)$ \\ \hline
Flatten & $(B,22)$ \\ \hline
Linear + GELU & $(B,22)$ \\ \hline
Linear + GELU & $(B,22)$ \\ \hline
Linear + GELU & $(B,22)$ \\ \hline
Linear + GELU & $(B,22)$ \\ \hline
Linear & $(B,1)$ \\ \hline
\end{tabular}%
}
\caption{The architecture of the encoder starting with the Fourier transform layer. The same decoder architecture is used as in Tab.~\ref{tab:cnn_ae}. $B$ indicates the batch size.}
\label{tab:fourier_encoder}
\end{table}

\section{Encoder output without scaling}

Due to the inherent randomness of neural network training, different models—even with identical architectures and training procedures—can produce different representations in the bottleneck layer as a function of scatterer separation. 

\begin{figure}
    \centering
    \includegraphics[width=\columnwidth]{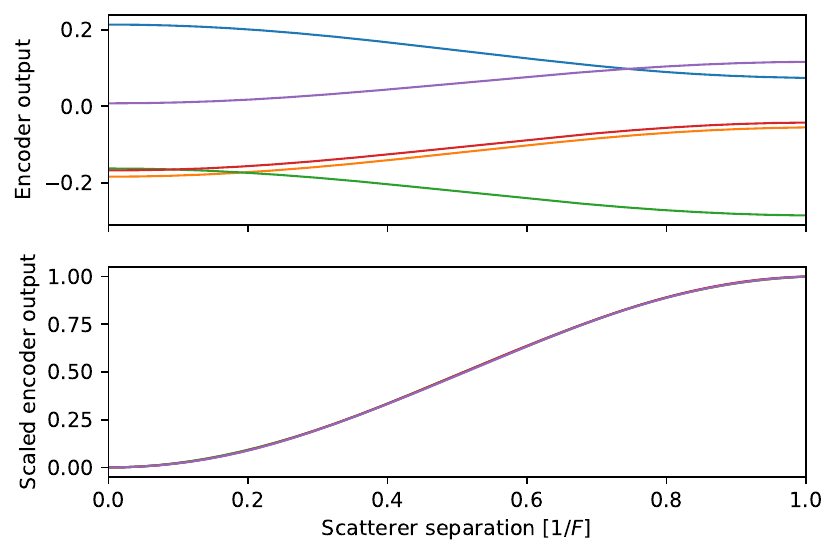}
    \caption{
    Encoder outputs from five selected neural networks with architecture defined in Table~\ref{tab:cnn_ae}, all trained on Bessel signals. The top panel shows the raw (unscaled) encoder outputs as a function of scatterer separation, while the bottom panel presents the same outputs scaled according to Eq.~(\ref{eq:clean_signal_scaling}). Each color corresponds to a different neural network and is consistent across both plots. The overlap of all curves in the bottom plot demonstrates the reproducibility and consistency achieved through scaling.
    }
    \label{fig:unscaled_output}    
\end{figure}

However, a reproducible structure in the encoder output emerges when scaling is applied. Fig.~\ref{fig:unscaled_output} illustrates the difference in bottleneck behavior with and without scaling, showing that the rescaling procedure enables meaningful comparisons across independently trained networks.

\section{Training of the autoencoders}

All models used for noiseless signal reconstruction are trained for $3000$ epochs with a batch size of 512. The batch consists of signals computed for with equally-spaced scatterer separations taken from range $[0,1]$. If trained on noisy signals, the number of epochs was increased to $5000$.
In each epoch, a random set of noise tensors is generated and added to the clean signals.
The training criterion is the mean squared error, and the optimization is performed using the Adam optimizer \cite{Adam} with a learning rate of 0.001, keeping the default PyTorch parameters.
\newline

\section{Dependence of the results on network architecture}\label{app:other_network_type_results}

The results presented in the main text are based on a convolutional autoencoder architecture, as outlined in Table~\ref{tab:cnn_ae}. In this appendix, we assess whether the key findings—specifically the reproducibility of bottleneck layer behavior—are robust to changes in the encoder architecture.

\begin{figure}
    \centering
    \includegraphics[width=\columnwidth]{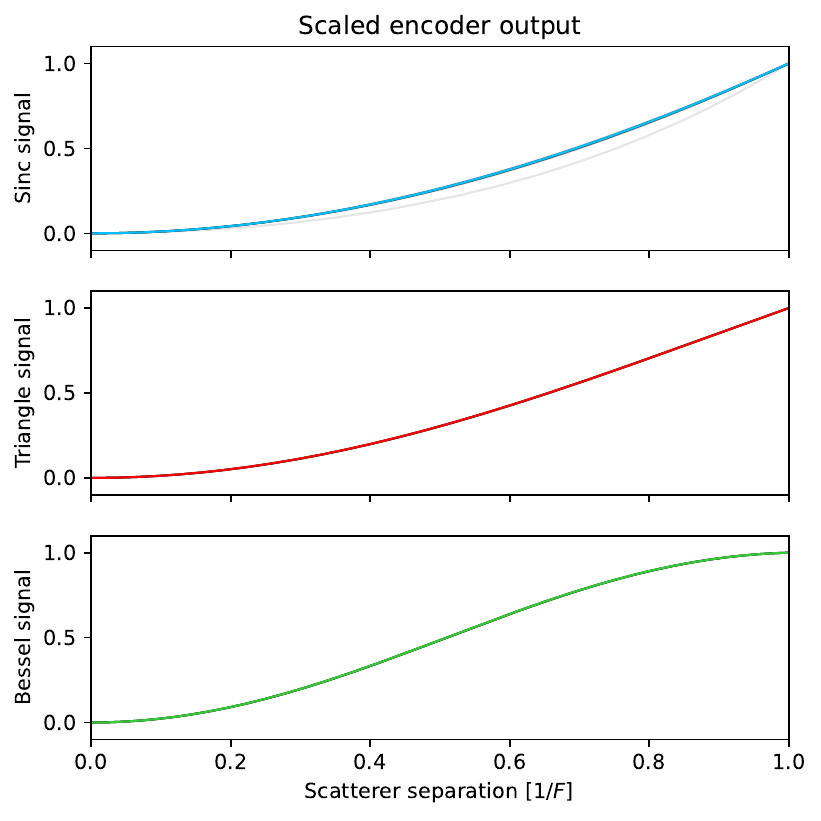}
    \caption{
    Encoder output scaled according to Eq.~(\ref{eq:clean_signal_scaling}) as a function of scatterer separation, plotted for the linear encoder architecture defined in Table~\ref{tab:linear_encoder}. Each panel corresponds to a different outgoing signal type. The 20 gray curves in each plot represent independently trained networks, and the colored curve (blue, red, or green) highlights the output of a representative model.
    }
    \label{fig:clean_linear_models}    
\end{figure}

\begin{figure}
    \centering
    \includegraphics[width=\columnwidth]{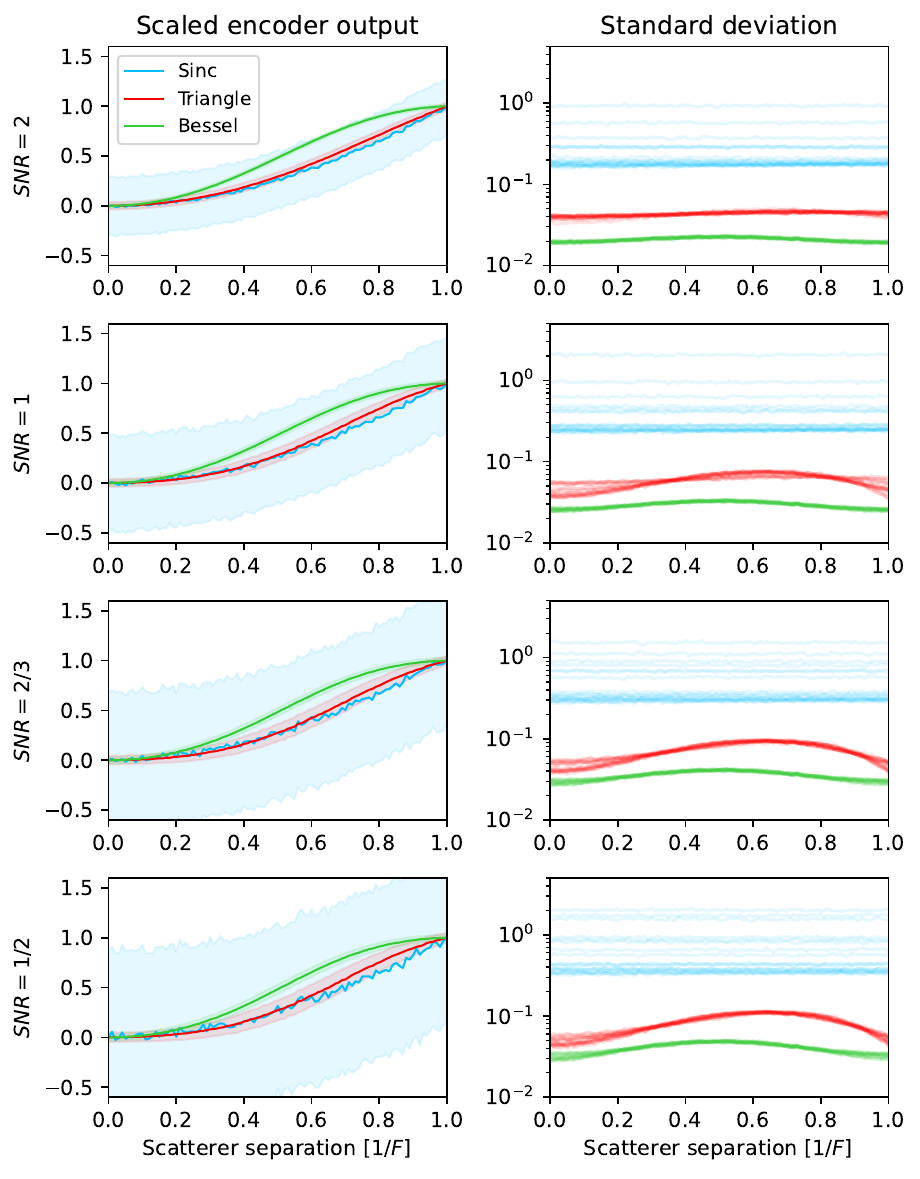}
    \caption{
    Results for the linear encoder architecture defined in Table~\ref{tab:linear_encoder}, plotted in the same format as Fig.~\ref{fig:noise_analysis}.  
    Left column: Scaled mean encoder output (solid lines) ± one standard deviation (shaded region) as a function of scatterer separation, shown for a representative model. Each color represents a different type of outgoing signal.  
    Right column: Standard deviation of the scaled encoder output plotted for all 20 independently trained networks, with each curve corresponding to a different training instance. Scaling was performed according to Eq.~(\ref{eq:noisy_signal_scaling}).  
    Each row corresponds to a different signal-to-noise ratio (SNR).
    }
    \label{fig:noise_analysis_linear_models}    
\end{figure}

We first examine a purely linear encoder, as detailed in Table~\ref{tab:linear_encoder}. The corresponding results are shown in Figures~\ref{fig:clean_linear_models}-\ref{fig:noise_analysis_linear_models}. For clean signals, the scaled encoder output exhibits the same monotonic and reproducible trends as observed in Fig.~\ref{fig:clean_signal_encoder_output}. 
For noisy signals, a hierarchy in performance persists: the Bessel-based signal consistently yields the lowest standard deviation in the bottleneck output, followed by the triangle and sinc signals, mirroring the behavior reported in Fig.~\ref{fig:noise_analysis}.

\begin{figure}
    \centering
    \includegraphics[width=\columnwidth]{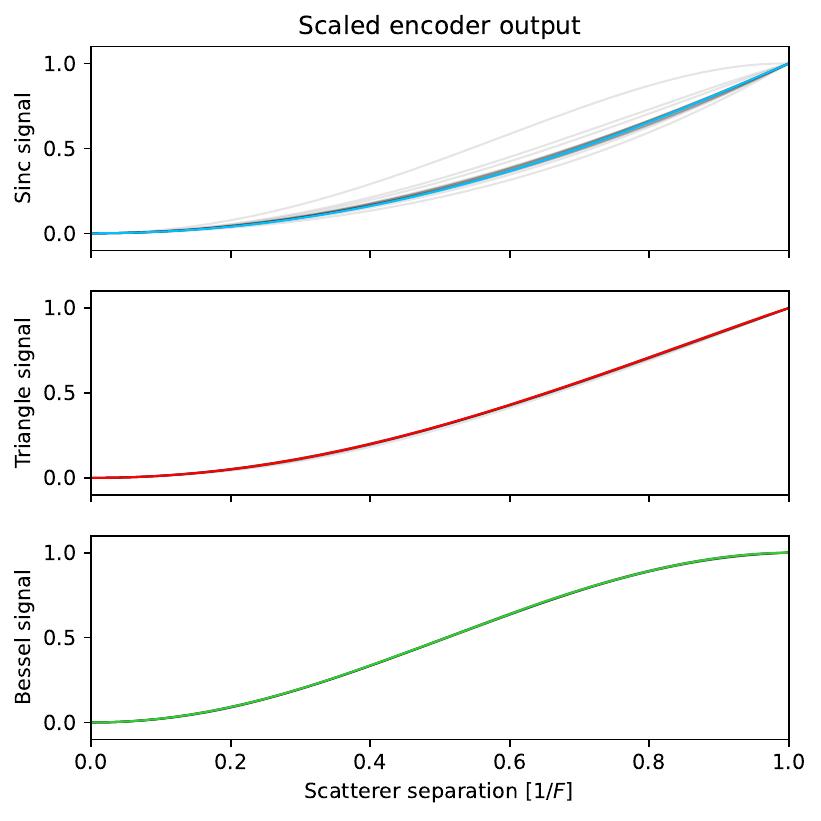}
    \caption{
    Encoder output scaled by Eq.~(\ref{eq:clean_signal_scaling}) as a function of scatterer separation, produced using the Fourier-based encoder architecture defined in Table~\ref{tab:fourier_encoder}. Each panel corresponds to a different outgoing signal type. The 20 gray curves in each plot represent outputs from independently trained networks, while the colored curve (blue, red, or green) highlights a representative model. 
    }
    \label{fig:clean_fourier_models}    
\end{figure}

\begin{figure}
    \centering
    \includegraphics[width=\columnwidth]{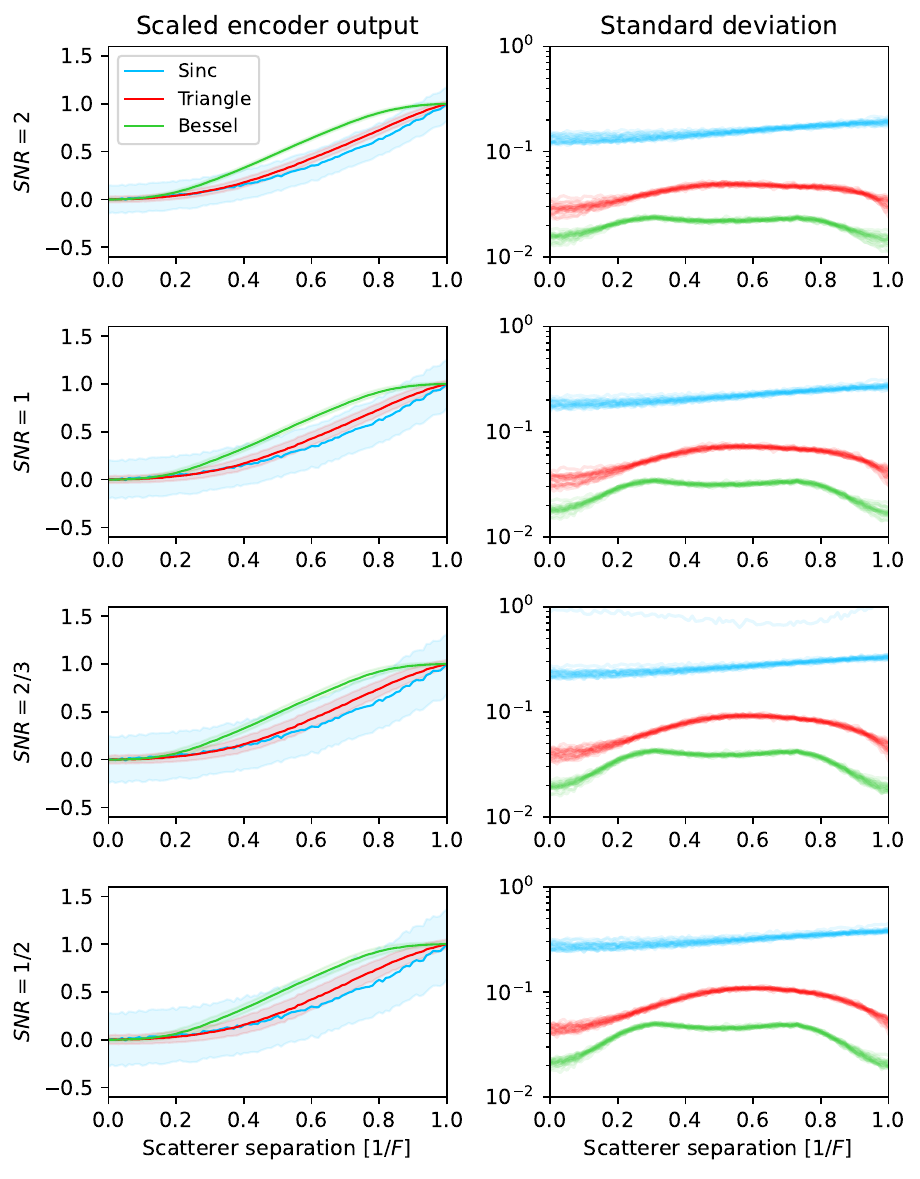}
    \caption{
    Results for the Fourier-based encoder architecture defined in Table~\ref{tab:fourier_encoder}, shown in the same format as Figure~\ref{fig:noise_analysis}.  
    Left column: Scaled mean encoder output (solid lines) ± one standard deviation (shaded region) as a function of scatterer separation, for a representative model. Each color corresponds to a different outgoing signal type.  
    Right column: Standard deviation of the scaled encoder output plotted for all $20$ independently trained networks per signal type. Scaling is performed according to Eq.~(\ref{eq:noisy_signal_scaling}).  
    Each row corresponds to a different signal-to-noise ratio (SNR). The observed trends—including the performance hierarchy (Bessel, triangle, sinc)—are consistent with results from the larger architectures, confirming the effectiveness of the Fourier-based encoder.
    }
    \label{fig:noise_analysis_fourier_models}    
\end{figure}

Next, we consider a more compact architecture based on a Fourier transform and low-pass filtering, as described in Table~\ref{tab:fourier_encoder}. The corresponding results are presented in Figures~\ref{fig:clean_fourier_models}-\ref{fig:noise_analysis_fourier_models}. A similar behavior of bottleneck layer is reproduced to the one reported for other architectures.

\end{document}